\documentclass[aps,prd,twocolumn,10pt,
tightenlines,superscriptaddress,
nofootinbib,showpacs]{revtex4-1}
\usepackage{amssymb,latexsym}
\usepackage{amsmath,amsbsy,bbm}
\usepackage{epsfig,bm}
\usepackage{graphicx,comment}
\usepackage{epstopdf}

\DeclareMathOperator{\tr}{tr}

\begin{document}

\def\ol#1{{\overline{#1}}}
\def\eqref#1{{(\ref{#1})}}

\def\Dslash{D\hskip-0.65em /}

\def\CPT{{$\chi$PT}}
\def\diag{\text{diag}}

\def\a{{\alpha}}
\def\b{{\beta}}
\def\d{{\delta}}
\def\D{{\Delta}}
\def\e{{\varepsilon}}
\def\g{{\gamma}}
\def\G{{\Gamma}}
\def\k{{\kappa}}
\def\l{{\lambda}}
\def\L{{\Lambda}}
\def\m{{\mu}}
\def\n{{\nu}}
\def\o{{\omega}}
\def\O{{\Omega}}
\def\S{{\Sigma}}
\def\s{{\sigma}}
\def\th{{\theta}}

\def\cF{{\mathcal F}}
\def\cS{{\mathcal S}}
\def\cC{{\mathcal C}}
\def\cE{{\mathcal E}}
\def\cB{{\mathcal B}}
\def\cT{{\mathcal T}}
\def\cQ{{\mathcal Q}}
\def\cL{{\mathcal L}}
\def\cO{{\mathcal O}}
\def\cA{{\mathcal A}}
\def\cV{{\mathcal V}}
\def\cR{{\mathcal R}}
\def\cH{{\mathcal H}}
\def\cW{{\mathcal W}}
\def\cM{{\mathcal M}}
\def\cD{{\mathcal D}}
\def\cN{{\mathcal N}}
\def\cP{{\mathcal P}}
\def\cK{{\mathcal K}}
\def\cU{{\mathcal U}}

\title{Low-Energy QCD in the Delta Regime}

\author{Matthew E.~Matzelle}
\email[]{Matt.Matzelle@gmail.com}
\affiliation{
Department of Physics,
        The City College of New York,
        New York, NY 10031, USA}
\author{Brian~C.~Tiburzi}
\email[]{btiburzi@ccny.cuny.edu}
\affiliation{
Department of Physics,
        The City College of New York,
        New York, NY 10031, USA}
\affiliation{
Graduate School and University Center,
        The City University of New York,
        New York, NY 10016, USA}
\affiliation{
RIKEN BNL Research Center,
        Brookhaven National Laboratory,
        Upton, NY 11973, USA}
\date{\today}

\date{\today}


\begin{abstract}

We investigate properties of low-energy QCD in a finite spatial volume, 
but with arbitrary temperature. 
In the limit of small temperature
and small cube size compared to the pion Compton wavelength, 
Leutwyler has shown that the effective theory describing low-energy QCD reduces to that of quantum mechanics on the coset manifold, 
which is the so-called delta regime of chiral perturbation theory.
We solve this quantum mechanics analytically for the case of a 
$U(1)_L \times U(1)_R$
subgroup of chiral symmetry, 
and numerically for the case of 
$SU(2)_L \times SU(2)_R$.
We utilize the quantum mechanical spectrum to compute the mass gap and chiral condensate, 
and investigate symmetry restoration in a finite spatial volume as a function of temperature. 
Because we obtain the spectrum for non-zero values of the quark mass, 
we are able to interpolate between the rigid rotor limit, 
which emerges at vanishing quark mass, 
and the harmonic approximation, 
which is referred to as the
$p$-regime.
We find that the applicability of perturbation theory about the rotor limit largely requires lighter-than-physical quarks. 
As a stringent check of our results, 
we raise the temperature to that of the inverse cube size. 
When this condition is met, 
the quantum mechanics reduces to a matrix model. 
The condensate we obtain in this limit agrees with that determined analytically in the epsilon regime.

\end{abstract}
\maketitle

\section{Introduction}

One of the hallmark features of low-energy QCD is spontaneous breaking of chiral symmetry. 
In the limit that the up and down quarks are massless, 
the QCD action has a
$U(2)_L \times U(2)_R$
symmetry, 
however, 
the 
$U(1)_A$
subgroup does not remain a symmetry at the quantum level. 
The remaining chiral symmetry, 
however,  
is hidden due to spontaneous symmetry breakdown to the 
vector isospin subgroup, 
$SU(2)_L \times SU(2)_R \to SU(2)_V$. 
While the up and down quarks are not massless in nature, 
their masses are considerably small compared to the QCD scale, 
and the pions can be identified as the pseudo-Goldstone bosons of the broken chiral symmetry. 
Further consequences of the Goldstone-boson character of pions are comprehensively detailed in%
~\cite{Donoghue:1992dd}, 
for example.

\begin{center}
\begin{figure}
\epsfig{file=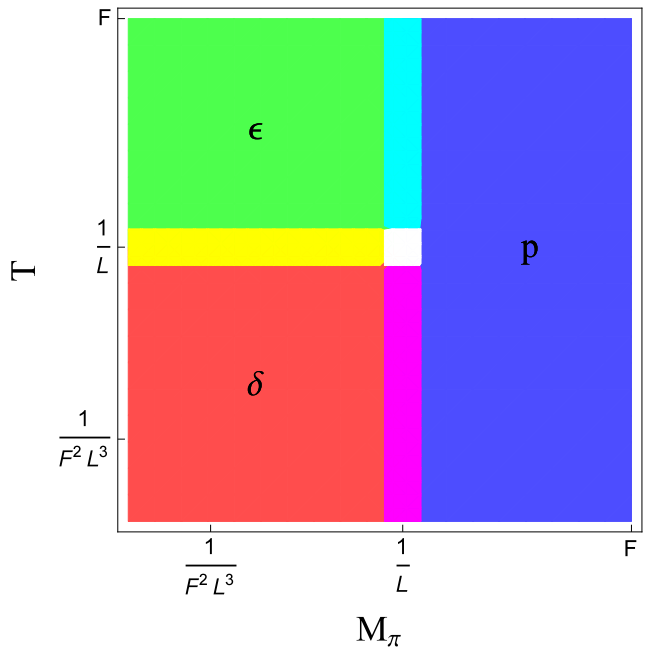,width=0.475\textwidth}
\caption{
Different regimes of the low-energy chiral expansion in finite volume. 
We label the regimes using schematic, color-coded regions of applicability that depend on temperature and pion mass
($\e=$ green, $\delta=$ red, $p=$ blue). 
We are specifically interested in the heart of the 
$\delta$
regime, 
and previous studies have been limited to the slice at 
$M_\pi = 0$, 
the region perturbatively close to this slice, 
or the region where 
$\d$- 
and 
$p$-regimes overlap. 
}
\label{f:realms}
\end{figure}
\end{center}

The fact that spontaneous chiral symmetry breaking occurs has now been rigorously established from first-principles lattice gauge theory computations, 
see%
~\cite{DeGrand:2006zz}
for an overview of lattice QCD methods. 
These computations are necessarily performed using a finite, Euclidean spacetime volume.  
In the absence of explicit chiral symmetry breaking introduced by the quark mass, 
the order parameter for chiral symmetry breaking, 
the so-called chiral condensate, 
remains zero in finite volume. 
With finitely many degrees of freedom, 
quantum tunneling becomes possible and the dynamics dictate that equivalent vacua are averaged over in a chirally symmetric fashion. 
This finite-volume restoration of chiral symmetry has 
been elucidated directly from the 
low-energy effective field theory%
~\cite{Gasser:1986vb}. 
The effective theory is chiral perturbation theory; 
and, 
in small, periodic spacetime volumes, 
the zero four-momentum mode of the pion becomes strongly coupled upsetting the infinite-volume power counting of the effective theory.  
Ultimately the non-perturbative dynamics of the zero mode leads to depletion of the chiral condensate, 
and this effect can be computed utilizing a modified power-counting scheme that appropriately treats the zero mode. 
This is referred to as the 
$\e$-regime 
of chiral perturbation theory, 
and its region of applicability is depicted in 
Fig.~\ref{f:realms}. 
Infinite volume properties of low-energy QCD remain accessible through correlation functions that are determined in the 
$\e$-regime, 
however, 
one must account for the non-trivial finite-volume effects from zero modes. 
For early work in this direction, 
see~\cite{Hasenfratz:1989pk,Hansen:1990un}.
This regime of finite volume QCD has received a lot of attention, 
in particular due to the relation to random-matrix theory 
and the spectrum of the Dirac operator%
~\cite{Verbaarschot:1994qf,Shuryak:1992pi}. 
For a detailed review, 
see%
~\cite{Verbaarschot:2000dy}.

With non-zero quark mass, 
and near the infinite volume limit, 
one enters the 
$p$-regime%
~\cite{Gasser:1987zq}, 
in which the standard power-counting of chiral perturbation theory applies, 
albeit with quantized momentum modes. 
Power-counting schemes for regimes intermediate to these two have also been proposed%
~\cite{Detmold:2004ap,Aoki:2011pza}.
In this work, 
we focus on a considerably less explored regime of finite-volume chiral perturbation theory. 
This is the novel
$\d$-regime,  
which was first explicated by Leutwyler%
~\cite{Leutwyler:1987ak}. 
It emerges when the pion Compton wavelength is larger than the spatial box size, 
for which the zero three-momentum mode becomes strongly coupled. 
For two massless light-quark flavors, 
the theory can be elegantly cast into a quantum mechanical Hamiltonian for an
$SO(4)$
rigid rotor.  
The spectrum exhibits a non-zero gap at vanishing quark mass, 
which accordingly demonstrates that chiral symmetry does not break at finite spatial volume. 
Corrections to the rotor spectrum have been determined using the next-to-leading order 
chiral Lagrangian in the chiral limit%
~\cite{Hasenfratz:2009mp,Niedermayer:2010mx}. 
The mass gap in the $\d$-regime was the focus of a lattice QCD computation,
although results were extrapolated from the 
$p$-regime%
~\cite{Bietenholz:2010az}.
Connections of the rotor limit of QCD to analogous condensed matter systems have also been described%
~\cite{Chandrasekharan:2006wn}. 
In the present note, 
our central concern is with extending the range of existing 
$\d$-regime 
results to non-zero values of the quark mass. 
We treat the quark mass according to the original 
$\d$-regime 
power counting, 
rather than additionally considering a perturbative quark mass expansion, 
see Fig.~\ref{f:realms}. 
In the case of two light-quark flavors in the 
$\d$-regime, 
the determined quark mass dependence allows us to investigate the rather rich behavior of the low-energy spectrum. 
Additionally we investigate the limitations of a perturbative treatment of the quark mass.

Our discussion is ordered in the following way. 
In 
Sec.~\ref{PC},  
salient features of low-energy QCD in the 
$\d$-regime 
are reviewed, 
including the 
$\d$-regime power counting. 
In Sec.~\ref{U1}, 
we treat the case of a 
$U(1)_L \times U(1)_R$ 
subgroup of chiral symmetry. 
We obtain the energy spectrum and partition function from eigenstates of the corresponding
$\d$-regime Hamiltonian, 
which is solvable in terms of the well-known Mathieu functions. 
The spectrum is utilized to determine the mass gap and chiral condensate, 
and their properties are explored as a function of quark mass and volume. 
Additionally the results are confirmed by matching up with the 
$\e$-regime. 
In Sec.~\ref{SU2}, 
the two-flavor case of chiral symmetry is treated. 
This section follows the evolution of the previous, 
however, 
eigenvalues of the Hamiltonian are determined numerically from matrix inversion, 
and results are checked against known limiting cases. 
Our findings are summarized in 
Sec.~\ref{summy}.

%
%
%
%
%
%
%
%
%
%

\section{Power counting in the delta regime} \label{PC}

Let us start with the effective low-energy representation of the QCD partition function,  
$Z = \tr (e^{-\b H_\text{QCD}})$, 
given by%
~\cite{Gasser:1983yg}
\begin{equation}
Z 
= 
\int \cD \cU
\,
e^{- S[\cU]}
,\end{equation}
where the low-energy action 
$S$ 
is written in terms of a Lagrangian density 
$\cL$
\begin{equation} \label{eq:S}
S 
= 
\int_0^\b dt \int_0^L d\bm{x} 
\, 
\cL(\bm{x}, t)
,\end{equation}
with 
$L$ 
as the length of each spatial dimension, 
and 
$\b = 1 / T$ 
as the length of the Euclidean time direction.
At leading order, the chiral Lagrangian density appears as
\begin{equation}
\cL 
=
\frac{F^2}{4}  
\,
\tr 
\left[
\partial_\mu \cU^\dagger \partial_\mu \cU
-
2 B \cM
(
\cU^\dagger + \cU
) 
\right]
,\end{equation}
with 
$\cM$ 
as the quark mass matrix, 
while
$F$
is the chiral limit value of the pion decay constant. 
The parameter 
$B$
is related to the chiral limit value of the quark condensate,%
\footnote{
While we have chosen to work with the total condensate rather than the condensate per flavor, 
the trivial factor of 
$N_f$
ultimately cancels in the condensate ratios presented throughout. 
} 
namely
$< \ol \psi \psi  > = - N_f B F^2$. 
The pion fields 
$\phi$ 
are 
$SU(N_f)$ 
matrices, 
and are contained in 
$\cU$ 
as
\begin{equation}
\cU = \exp ( i \sqrt{2} \phi / F)
.\end{equation}
To completely define the theory in Eq.~\eqref{eq:S}, 
we take the pion fields 
$\phi$ 
to satisfy periodic boundary conditions in both space and time.
This will be the case for lattice simulations of QCD wherein the quark 
fields satisfy either periodic or antiperiodic boundary conditions in 
space and time. 
Consequently the allowed pion four-momenta, 
$p_\mu$, 
are quantized in modes, 
$n_\mu$, 
in the form
\begin{equation}
p_\mu 
= 
\left( 
\frac{2 \pi}{L} \bm{n}  
\,
,
\, 
2 \pi 
\, 
T 
\, 
n_4
\right)
,\end{equation}
where 
$n_\mu = (\bm{n}, n_4)$ 
is a four-vector of integers. 
Expanding the Lagrangian density to quadratic order in the pion fields, 
we see
$M_\pi^2 = 2 B m_q$
assuming 
$N_f$
degenerate quark flavors of mass 
$m_q$.

In this finite spacetime volume, 
a novel regime of low-energy QCD was pointed out by 
Leutwyler~\cite{Leutwyler:1987ak}.
Let $\d$ be a small parameter, 
and treat
\begin{equation} \label{eq:del}
M_\pi \sim T \sim \d^3
\quad
\text{and}
\quad
\frac{1}{L} \sim \delta
,\end{equation}
so that the pion Compton wavelength is larger than the box size,
$M_\pi L \sim \d^2$, 
but the small temperature $T$ is the same order as the pion mass,
$M_\pi / T  = M_\pi \beta \sim 1$. 
The pion propagator for the mode 
$n_\mu$ 
has the behavior
\begin{equation}
G(n_\mu) =
\left[
\left(
\frac{2 \pi \bm{n}}{L}
\right)^2 
+
\left( 
2 \pi T n_4
\right)^2 
+
M_\pi^2
\right]^{-1}
.\end{equation}
Propagation of modes with 
$\bm{n} \neq \bm{0}$ 
scale with 
$\d^{-2}$, 
while spatial zero modes, i.e.~those with 
$\bm{n} = \bm{0}$, 
scale with 
$\d^{-6}$.
Vertices from the leading-order chiral Lagrangian scale differently depending
on the momentum. A quark mass insertion or two temporal derivatives both
scale as 
$M_\pi^2$, 
$p_0^2 \sim \d^6$, 
while a spatial gradient vertex behaves like
$\bm{p}^2 \sim \d^2$. 
Loop diagrams come weighted with a spactime volume factor of
$(\b L^3)^{-1} \sim \d^6$.

A typical Feynman diagram having $\ell$ loops, $I$ internal lines, and $V$ vertices
has a counting depending on which modes are spatial zero modes and which are not. 
For all spatial zero modes propagating in the diagram, we have
\begin{equation}
\bm{n} = \bm{0} :
\qquad
\d^{6 (\ell - I + V)}
= 
\d^6
,\end{equation}
whereas for all non-zero modes, we have
\begin{equation} \label{eq:pcNzero}
\bm{n} \neq \bm{0} : 
\qquad
\d^{6 \ell - 2 I + 2 V}
= 
\d^6 \d^{4 (\ell - 1)}
.\end{equation}
For the non-zero modes, there is thus a loop expansion.%
\footnote{
In general there are Feynman diagrams with both zero 
and non-zero modes propagating. 
These will appear in the power counting 
between the two extremes detailed above. 
As we work to leading order in the 
$\d$-regime, 
such contributions are beyond our consideration.
} %
On the contrary, there is no loop expansion for the spatial zero modes, 
and we must work non-perturbatively in the leading-order chiral Lagrangian. 
To accomplish this, 
we use the collective variable 
$U(t)$ 
defined through the relation
\begin{equation} \label{eq:mode}
\cU(x) 
= 
\sqrt{U(t)}
\exp 
\left[ 
i \sqrt{2} \varphi(x) / F
\right]
\sqrt{U(t)}
,\end{equation}
where
\begin{equation}
\varphi(x) = 
\sum_{\bm{n} \neq \bm{0}, n_4} e^{i p \cdot x } 
\, \tilde{\varphi}_{n_\mu}
\end{equation}
has been defined to exclude the spatial zero modes of the pion fields. 
With the parametrization in Eq.~\eqref{eq:mode}, 
the chiral action for spatial zero modes becomes
\begin{eqnarray}
S 
&=& 
\frac{F^2 L^3}{4} 
\int_0^\b dt
\, \, 
\tr
\left[
\frac{\partial U^\dagger}{\partial t}
\frac{\partial U}{\partial t}
- 
2 B \cM
\left(
U {}^\dagger
+ 
U
\right)
\right]
,\quad 
\end{eqnarray}
up to corrections of order 
$\d^2$.

As anticipated by the power counting in Eq.~\eqref{eq:pcNzero}, 
there are one-loop contributions from the functional integral over the non-zero modes, 
however, 
the leading contribution to the QCD partition function is a mass-independent multiplicative factor. 
Thus in the 
$\d$-regime 
we have
\begin{equation}
Z = Z'  \,  \tr \left( e^{ - \b  H }   \right)
,\end{equation}
to 
$\cO(\delta^2)$, 
where the effective Hamiltonian is given by
\begin{equation}
H
= 
- 
\frac{1}{2 F^2 L^3} 
\mathfrak{D}^2
- 
\frac{1}{2}
B F^2 L^3 
\,  
\tr \left[ \cM \left( U^\dagger +  U \right) \right] 
,\end{equation} 
with 
$\mathfrak{D}^2$
as the Laplace-Beltrami operator on 
$SU(N_f)$. 
The non-zero mode path integration contributes
to the normalization factor 
$Z'$.
Any contributions arising from the ambiguity of 
operator ordering
have additionally been absorbed into this factor, 
which is possible because the curvature scalar of 
$SU(N_f)$ 
is constant. 
The constant
$Z'$ 
can be determined from 
matching to expressions calculated in the $p$-regime.
Up to the overall irrelevant normalization factor, 
we have
\begin{equation}
Z' = e^{- \b E_L}
,\end{equation}
where 
$E_L$ 
is the contribution to the vacuum energy, 
and is given by~\cite{Leutwyler:1987ak} 
\begin{equation}
E_L 
= 
\frac{N_f^2 -1}{2L}
\left[- 
\gamma_0
+ 
\frac{N_f }{(2  F L)^2}
\right]
,\end{equation}
for the case of $N_f$ quark flavors. 
Here $\gamma_0$ is a pure number characteristic of a spatial torus, 
and is given by
\begin{equation}
\gamma_0 = \frac{1}{\pi^2} \sum_{\bm{n} \neq \bm{0}} \frac{1}{\bm{n}^4}
.\end{equation}
The term in the vacuum energy involving
$\gamma_0$
can be interpreted as arising from the Casimir effect, 
while the small repulsive contribution is a perturbative correction due to non-zero mode scattering in the vacuum.

For ease below, 
we work with the dimensionless variables
\begin{eqnarray}
\tau 
&=&
2 F^2 L^3 T, 
\quad 
\text{and}
\quad
\mu = M_\pi F^2 L^3
.\end{eqnarray}
In terms of these variables, 
the partition function can be written as
\begin{equation}
Z
=
Z'
\tr
\left(
e^{- \cH / \tau}
\right)
,\end{equation}
where the dimensionless Hamiltonian is simply
\begin{equation} \label{eq:Ham}
\cH
= 
- 
\mathfrak{D}^2
- 
\frac{\mu^2}{2}
\tr \left( U^\dagger + U \right)
.\end{equation}
In what follows, 
we determine the spectrum of the Hamiltonian operator
$\cH$, 
which is the spectrum in the 
$\d$-regime up to the mass-independent shift
$E_L$. 
Of further consideration is the volume dependence of the chiral condensate, 
$< \ol \psi \psi >_L$. 
More precisely, 
we focus on the ratio of the finite volume condensate to that in infinite volume
\begin{equation}
\Sigma
= 
< \ol \psi \psi >_L 
/
< \ol \psi \psi >
,\end{equation} 
because it is QCD renormalization scale and scheme independent. 
Notice we treat the temperature dependence of this quantity as implicit. 
The condensate is determined from the derivative of the partition function logarithm
\begin{equation} 
< \ol \psi \psi >_L
= 
- 
\frac{1}{\beta L^3}
\frac{\partial \log Z}{\partial m_q}
,\end{equation}
from which we see
\begin{equation} \label{eq:chi}
\Sigma
=
\frac{\tau}{N_f}
\frac{\partial \log Z}{\partial \mu^2}
.\end{equation}
In this work, 
we restrict our attention to the two-flavor case,
$N_f = 2$.

\section{$U(1)_L \times U(1)_R$} \label{U1}

As the spectrum in the 
$\d$-regime 
for the two-flavor case ultimately requires a numerical approach, 
we begin by considering a simpler case which we show is solvable in terms of special functions. 
This case is that of an unbroken 
$U(1)_L \times U(1)_R$
subgroup of chiral symmetry, 
which in infinite volume is broken down to 
$U(1)_V$
by the formation of the chiral condensate. 
This symmetry breaking pattern can be realized in several ways: 
for example, 
the theory of QCD + QED with two massless quarks has this symmetry breaking pattern, 
for which the sole Goldstone boson is the neutral pion; 
an analogous situation occurs in QCD with an isospin chemical potential%
~\cite{Son:2000xc}, 
or that of QCD with isospin twisted boundary conditions%
~\cite{Mehen:2005fw}. 
Staggered fermions%
~\cite{Susskind:1976jm} 
on a coarse lattice present an additional example. 
In this case, 
the unbroken 
$U(1)_L \times U(1)_R$
subgroup at finite lattice spacing is contained in the larger chiral symmetry group of so-called quark taste, 
$SU(4)_L \times SU(4)_R$, 
see, 
e.g.%
~\cite{Golterman:1984cy,Lee:1999zxa}.

\subsection{Determination of the Spectrum}

For a theory with residual 
$U(1)_L \times U(1)_R$ 
chiral symmetry, 
the
Goldstone manifold is parametrized by a single angle $\a$, 
which we can choose to correspond to the neutral pion. 
To determine the spectrum in the 
$\d$-regime, 
we must consider the spatial zero mode, 
which thus has the form
\begin{equation}
U(t) 
= 
\exp{
[ 
i 
\alpha(t)
\tau^3
]
}
.\end{equation}
To solve the theory, 
we obtain the eigenvalues of the corresponding 
$\d$-regime 
Hamiltonian, 
which in dimensionless units reads
\begin{equation}
\cH 
= 
- 
\frac{d^2}{d \alpha^2}
- 
2\mu^2 
\cos \a
.\end{equation}
The corresponding eigenfunctions and eigenvalues are given by the well-known Mathieu functions and Mathieu characteristics, 
respectively. 
Periodicity in the parametrization of 
$U(1)$, 
namely 
$\alpha \equiv \alpha + 2 \pi$, 
restricts us to the 
$\pi$-periodic solutions of the Mathieu equation. 
These solutions can be further classified by their behavior under reflection
$\alpha \to - \alpha$. 
This corresponds directly to the parity transformation of QCD. 
Even and odd solutions are written as the Mathieu functions
$\texttt{ce}_{2 n}$,  
and 
$\texttt{se}_{2 n}$, 
respectively. 
The required eigenfunctions 
of the Hamiltonian 
$\cH$
are given by 
\begin{eqnarray}
\Psi_{n}^{(e)} (\alpha) 
&=&
N \texttt{ce}_{2n} \left( \frac{\alpha}{2},-4\mu^2 \right),
\notag \\
\Psi_{n}^{(o)}
(\a)
&=&
N \texttt{se}_{2n}
\left(
\frac{\a}{2},-4\mu^2
\right)
,\end{eqnarray}
where 
$n = 0$, $1$, $2$, $\cdots$
for the even solutions, 
and 
$n = 1$, $2$, $\cdots$
for the odd. 
The multiplicative factor of 
$N$
is determined by the wavefunction normalization; 
and, 
although we do not write it explicitly,  
the normalization generally depends on the parameters
$n$ 
and 
$\mu^2$. 
Notice that 
$\pi$-periodicity of the Mathieu functions implies the condition
$\Psi(\alpha + 2 \pi) = \Psi(\alpha)$.
The energy eigenvalues are given by 
\begin{eqnarray} \label{eq:energ}
E_{n}^{(e)}
&=&
\frac{1}{4} 
\texttt{a}_{2n}(-4\mu^2)
,\notag\\
E_{n}^{(o)}
&=&
\frac{1}{4}
\texttt{b}_{2n}(-4\mu^2)
,\end{eqnarray}
for the corresponding even and odd solutions. 
These are written in terms of 
$\texttt{a}_{2n}$
and
$\texttt{b}_{2n}$
which are referred to as Mathieu characteristics.

\begin{figure}
\epsfig{file=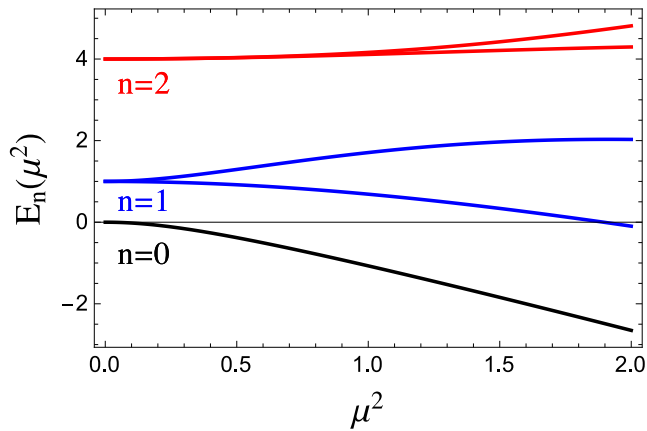,width=0.475\textwidth}
\caption{
Low-lying spectrum as a function of the parameter
$\mu^2 = (M_\pi L)^2 (F L)^4$
in the 
$\d$-regime of 
$U(1)_L \times U(1)_R$
chiral perturbation theory. 
The ground-state (vacuum) energy is shown, 
$n = 0$, 
as well as the energies of even- and odd-parity states with 
$n = 1$, $2$. 
}
\label{f:specU1}
\end{figure}

The low-lying spectrum of 
$\cH$
is shown as a function of 
$\mu^2$
in 
Fig.~\ref{f:specU1}.
The ground state (vacuum) is non-degenerate, 
while the even and odd solutions for a given 
$n$
are degenerate at 
$\mu^2 = 0$. 
Away from this value, 
the levels split, 
with a splitting that increases monotonically with 
$\mu^2$. 
Higher-lying states exhibit successively smaller splittings at fixed 
$\mu^2$, 
and hence parity doubling occurs. 
This is a manifestation of chiral symmetry restoration in the excited-state spectrum.

%
%
%
%
%
%
%
%
%
%

\subsection{Mass Gap}

\begin{figure}
\epsfig{file=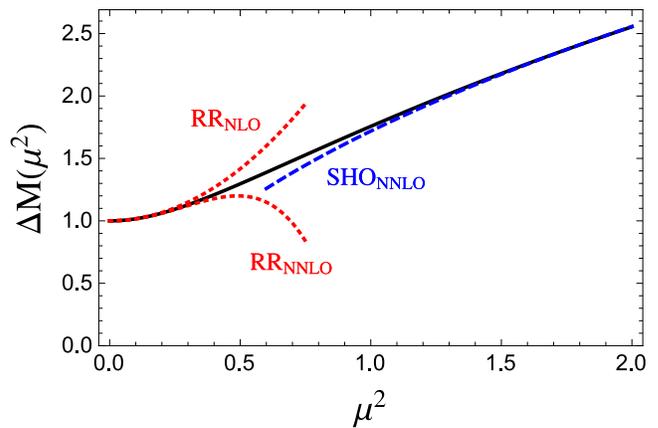,width=0.475\textwidth}
\caption{
Mass gap in the 
$\d$-regime of 
$U(1)_L \times U(1)_R$.
The gap, 
$\Delta M$
given 
in 
Eq.~\eqref{eq:DMU1},
is plotted as a function of 
$\mu^2$. 
Superimposed are results for the mass gap computed in perturbation theory for small 
$\mu^2$,
the rigid rotor 
(RR)
limit, 
and large 
$\mu^2$, 
the simple harmonic oscillator 
(SHO) 
limit.
For the SHO limit, 
the NLO and NNLO results are quite similar, 
and we have opted to plot the (slightly better) NNLO result. 
}
\label{f:gapU1}
\end{figure}

A simple quantity to investigate is the mass gap. 
From the spectrum, 
we see that the first excitation above the vacuum has odd parity and thus the quantum numbers of the neutral pion.
The mass gap 
$\Delta M$
is given by 
\begin{equation}
\D M
=
\frac{1}{4} 
\left[
\texttt{b}_2 ( - 4 \mu^2)
- 
\texttt{a}_0 (- 4 \mu^2)
\right]
\label{eq:DMU1}
.\end{equation}
At vanishing quark mass, 
$\mu = 0$, 
the mass gap is non-vanishing because chiral symmetry is restored in finite volume
excluding the possibility of a Goldstone pion. 
Increasing
$\mu^2$ 
away from zero introduces explicit chiral symmetry breaking, 
thus increasing the gap. 
This behavior is exhibited in Fig.~\ref{f:gapU1}, 
where we also compare the behavior of the mass gap computed in two limits using standard 
Rayleigh-Schr\"odinger perturbation theory. 
\begin{itemize}
\item
Rigid Rotor (RR).
At vanishing 
$\mu^2$, 
the effective Hamiltonian 
$\cH$
is that of a 
$U(1)$
rigid rotor, 
with eigenvalues 
$n^2$
and wavefunctions 
$\psi_n (\alpha) = \frac{1}{\sqrt{2 \pi}} e^{ i n \alpha}$, 
for 
$n = 0$, 
$\pm 1$, 
$\cdots$.
Treating the quark mass in perturbation theory leads to the mass gap
\begin{equation}
\D M^\text{RR}
=
1
+
\frac{5}{3}
\mu^4 
-
\frac{751}{216}
\mu^8
+ 
\cO(\mu^{12})
,\end{equation}
where the 
$\mu^4$
term we deem next-to-leading order 
(NLO)
and arises in second-order perturbation theory, 
while the 
$\mu^8$
term we deem next-to-next-to-leading order
(NNLO)
and arises in fourth-order perturbation theory. 
\item
Simple Harmonic Oscillator (SHO). 
In the opposite limit of large 
$\mu^2$, 
the potential term in 
$\cH$
forces the angle 
$\alpha$ near zero about which a harmonic approximation emerges.
In dimensionful units, 
the oscillator frequency
$\omega$ 
can be identified with 
$M_\pi$,
as one expects from quantum field theory of a zero-momentum mode.  
Perturbation theory can be performed about the harmonic limit, 
and leads to the mass gap
\begin{equation}
\D M^\text{SHO}
=
2 \mu
- 
\frac{5}{16}
-
\frac{9}{256}
\mu^{-1}
+ 
\cO(\mu^{-2})
.\end{equation}
The first term is merely the difference in oscillator quanta, 
whereas the second term arises in first-order perturbation theory
(NLO), 
and the third term arises at second order 
(NNLO). 
Reinstating dimensionful parameters, 
we appropriately recover the pion mass from the mass gap in the infinite volume limit
$\frac{1}{2 F^2 L^3} \Delta M^{\text{SHO}}  \overset{L \to \infty}{=} M_\pi$. 

\end{itemize}

%
%
%
%
%
%
%
%
%
%

From Fig.~\ref{f:gapU1}, 
we see that the 
SHO
approximation appears to work for the mass gap down to 
$\mu^2 \sim 1$, 
which is quite fortuitous given that the harmonic approximation emerges from considering
$\mu \gg 1$. 
One should keep in mind that for values of 
$M_\pi L \gg 1$, 
the non-zero modes are no longer suppressed, 
and these must be accounted for to recover the full results of the 
$p$-regime.  
Our harmonic approximation treats only the zero-momentum mode.

In the opposite limit,
which is the chiral limit,  
we see that the 
RR
approximation for the mass gap remains under perturbative control up to values of 
$\mu^2 \lesssim \frac{1}{2}$. 
In order for chiral expansion itself to be valid, 
we require
$(2 F L)^2 \gg 1$. 
Assuming that 
$F L > 1$
is sufficient leads to the requirement that 
$L > 2.2 \, \texttt{fm}$.   
In turn, 
requiring the RR approximation to be under control requires
$M_\pi L \gtrsim 1 / \sqrt{2}$. 
Combining the two constraints leads to a requirement on the pion mass, 
namely
$M_\pi \lesssim  65 \, \texttt{MeV}$. 
While this condition is quite prohibitive for lattice QCD computations, 
it can be avoided by treating the quark mass non-perturbatively, 
as we have done in Eq.~\eqref{eq:DMU1}.

%
%
%
%
%
%
\begin{figure}
\epsfig{file=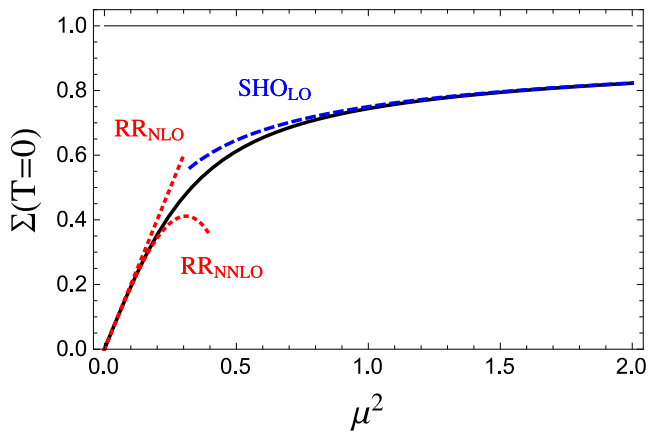,width=0.475\textwidth}
\caption{
Chiral condensate in the 
$\d$-regime 
of 
$U(1)_L \times U(1)_R$. 
Shown is the ratio of the chiral condensate at finite volume to that in infinite volume at vanishing temperature,
$T = 0$. 
Results obtained from perturbing about the rigid rotor limit
(RR) 
are shown along with the simple harmonic oscillator approximation
(SHO). 
}
\label{f:sigU1}
\end{figure}
%
%
%
%

\subsection{Chiral Condensate}

The chiral condensate is a quantity of particular interest in the 
$\d$-regime. 
At zero temperature, 
the chiral condensate can be determined solely from the quark mass derivative of the ground-state energy eigenvalue. 
In particular, 
we have the condensate ratio of finite to infinite volume given simply by the expression
\begin{equation}
\Sigma(T=0)
=
- \frac{1}{8} 
\frac{\partial
\texttt{a}_0 ( - 4 \mu^2)}{\partial \mu^2}
.\end{equation}
The behavior of this ratio is shown in Fig.~\ref{f:sigU1}.
In the chiral limit, 
$\mu^2 = 0$, 
the condensate identically vanishes
because there is no spontaneous chiral symmetry breaking at finite volume. 
A non-zero value of the quark mass introduces explicit breaking of chiral symmetry, 
and a bias for the vacuum expectation value of the coset field to lie near unity, 
corresponding to 
$\alpha = 0$. 
Consequently the condensate increases rapidly away from 
$\mu^2 = 0$. 
Also shown in the figure are perturbative approximations to the chiral condensate about the RR limit. 
Up to NNLO accuracy, 
we have
\begin{equation}
\Sigma^\text{RR}(T=0)
=
2 \mu^2 
- 
7 \mu^6
+ 
\cO(\mu^{10})
.\end{equation}
Unlike the mass gap (and the energy eigenvalues quite generally), 
the condensate requires non-zero 
$\mu^2$
in order not to vanish. 
As a consequence, 
the expansion of the chiral condensate about the RR limit does not perform as well as compared to that for the mass gap. 
From the figure, 
the expansion appears to be under control up to 
$\mu^2 \lesssim \frac{1}{4}$, 
which corresponds to pion masses 
$M_\pi \lesssim 45 \, \texttt{MeV}$
on the minimally sufficient volume of 
$L = 2.2 \, \texttt{fm}$.

Approaching from the opposite limit, 
$M_\pi L \gtrsim 1$,  
the SHO approximation for the chiral condensate works remarkably well.
Using only the energy eigenvalues of the simple harmonic oscillator without any perturbative corrections, 
we have the behavior of the zero-temperature condensate as
\begin{equation}
\Sigma^\text{SHO}(T=0)
=
1
- 
\frac{1}{4} 
\mu^{-1}
+ 
\cO(\mu^{-2})
,\end{equation}
which produces the LO curve shown in the figure. 
While the SHO result approaches the infinite volume limit slowly, 
i.e.~only with a power of $L$, 
namely the volume 
$L^{-3}$, 
we must remember that for 
$M_\pi L \gg 1$
contributions from non-zero modes become important.
Their summation results in exponential scaling, 
$\propto e^{ - M_\pi L } / (M_\pi L)^{3/2}$, 
rather than power law.

%
%
%
%
%
%
\begin{figure}
\epsfig{file=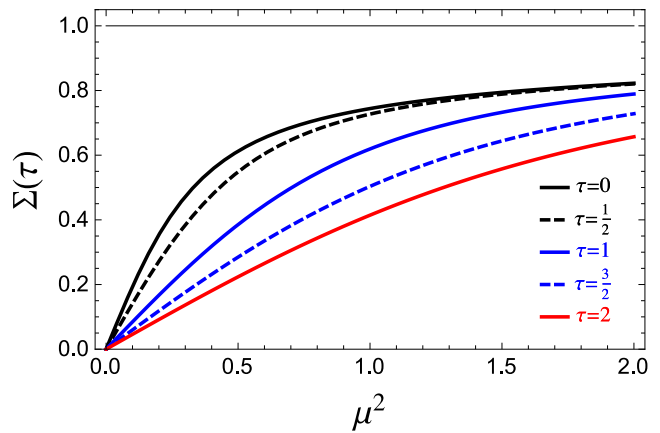,width=0.475\textwidth}
\caption{
Chiral condensate in the 
$\d$-regime 
of 
$U(1)_L \times U(1)_R$
as a function of mass 
$\mu^2$
and temperature. 
Shown is the ratio of the chiral condensate at finite volume to that in infinite volume for several values of the temperature, 
$\tau = 2 F^2 L^3 T$.
\label{f:sigTU1}
}
\end{figure}
%
%
%
%

As we increase the temperature away from zero, 
we require excited-state contributions to the partition function, 
which takes the form
\begin{equation}
Z
=
e^{- \frac{\texttt{a}_0 (- 4 \mu^2)}{4 \tau} }
+
\sum_{n=1}^{\infty}
\left[
e^{- \frac{\texttt{a}_{2n} (- 4 \mu^2)}{4 \tau}}
+
e^{- \frac{\texttt{b}_{2n} (- 4 \mu^2)}{4 \tau}}
\right] \label{eq:ZU1}
.\end{equation}
The temperature dependence of the chiral condensate ratio determined using the partition function is exhibited in 
Fig.~\ref{f:sigTU1}. 
Increasing the temperature not surprisingly melts the condensate. 
To determine the condensate, 
we truncate the infinite sum over energy levels. 
For the range of temperatures considered, 
$0 \leq \tau \leq 2$, 
a total of 
$13$
states, 
corresponding to even- and odd-parity states with  
$n \leq 6$,
is more than sufficient to guarantee convergence of the condensate. 
The maximum temperature plotted,
$\tau = 2$,
corresponds to 
$T = 90 \, \texttt{MeV}$
in physical units
for the minimal box size, 
$L = 2.2 \, \texttt{fm}$.

\begin{figure}
\epsfig{file=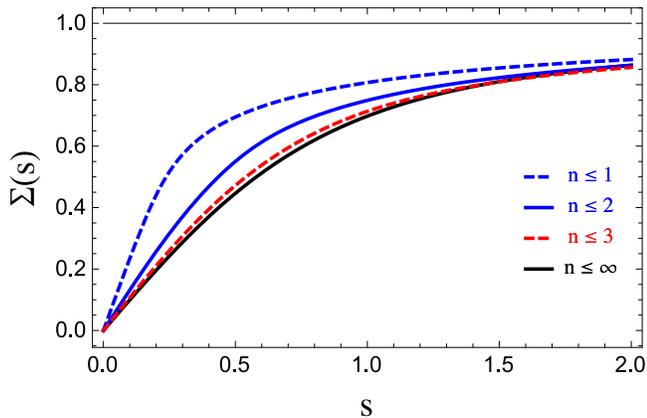,width=0.475\textwidth}
\caption{
Chiral condensate in the 
$\e$-regime 
of 
$U(1)_L \times U(1)_R$.
Also shown is the convergence of the 
$\d$-regime condensate with the quantum number 
$n$
(including both even- and odd-parity states for each 
$n$). 
In order to make this comparison, 
we set 
$F L  = 2$
corresponding to a temperature of
$T = 45 \, \texttt{MeV}$. 
With 
$n \leq 4$, 
the condensate ratio is indistinguishable from that determined directly in the 
$\e$-regime, 
which we label with 
$n \leq \infty$. 
\label{f:sigepsU1}
}
\end{figure}

Finally we increase the temperature to satisfy the condition 
$T = L^{-1}$, 
so that now
$M_\pi \beta \ll 1$. 
Under this condition, 
the $\d$-regime condensate can be verified against the known 
result from the 
$\e$-regime. 
In the latter regime, 
the pion fluctuations are frozen into the zero four-momentum mode, 
$n_\mu = (0,0,0,0)$, 
and there is no ``momentum'' on the group manifold driving the selection of the ground state. 
Instead, 
all values of 
$\a$
are averaged over with a weight that depends on the potential. 
The trace required for the partition function is then to be evaluated in the group-coordinate basis
\begin{equation}
Z 
= 
\tr \left( e^{ -  \b H} \right)
=
\frac{1}{2\pi} \int_0^{2 \pi} d \a \, e^{ 2 s \cos \a}
,\end{equation}
where we have introduced the variable 
$s = \mu^2 / \tau$, 
which satisfies
$s = \frac {1}{2} (M_\pi L)^2 (F L)^2$, 
for the particular value
$T = L^{-1}$. 
From this 
$\varepsilon$-regime partition function, 
one easily derives the finite volume to infinite volume ratio of the chiral condensate
\begin{equation}
\Sigma (s)
= 
\texttt{I}_1 (2s) / \texttt{I}_0 (2 s)
,\end{equation}
where the
$\texttt{I}_n (x)$ 
are modified Bessel functions. 
To compare this with the 
$\d$-regime result, 
we evaluate the latter by writing the mass in terms of the scaling variable $s$, 
namely 
$\mu^2 = s \tau = 2 s  (F L)^2$. 
For the convergence of the chiral expansion, 
we need 
$F L \gtrsim 1$; 
and, 
in practice, 
we choose the value
$F L = 2$ 
to match the 
$\d$- and $\e$-regime results.
This corresponds to a temperature of 
$T = 45 \, \texttt{MeV}$, 
where only a few excited states are required in the partition function, 
Eq.~\eqref{eq:ZU1}. 
Convergence with the number of states is demonstrated in 
Fig.~\ref{f:sigepsU1}. 
Notice the increased allowance of contributions from higher-lying states serves to melt the condensate, 
as chiral symmetry becomes restored in the spectrum of excited states. 
Having examined low-energy QCD in the $\d$-regime of 
$U(1)_L \times U(1)_R$, 
we now turn to the case of 
$SU(2)_L \times SU(2)_R$.

%
%
%
%
%
%
%
%
%
%

\section{$SU(2)_L\times SU(2)_R$} 
\label{SU2}

We begin our investigation of low-energy QCD in the $\delta$-regime of 
$SU(2)_L\times SU(2)_R$ 
with a brief discussion of our parametrization of the zero-mode manifold, 
and the resulting Hamiltonian. 
Next we solve for the eigenstates of zero isospin,  
$\ell=0$, 
for which there is an exact solution. 
This solution enables us to analytically determine the finite volume modification of the chiral condensate at zero temperature.
To obtain the spectrum of states with non-vanishing isospin, 
$\ell \neq 0$,  
we use a numerical method, 
for which comparison with the exact 
$\ell = 0$
result proves beneficial. 
As in Sec.~\ref{U1}, 
we calculate the mass gap, 
the finite volume modification to the chiral condensate at vanishing and non-vanishing temperature, 
and finally we show our results for the condensate appropriately match onto those determined in the 
$\e$-regime.

\subsection{Setup}

For the case of $SU(2)_L\times SU(2)_R$  symmetry
in the $\d$-regime, 
the spatial zero mode of the coset manifold must be treated non-perturbatively. 
Accordingly we adopt the standard parametrization  
\begin{equation}
U(t) = 
\exp 
\left[
i 
\alpha(t)
\,
\hat{\bm{n}}(t) \cdot \bm{\tau}
\right] 
\label{eq:coset2}
,\end{equation}
written in terms of the angle
$\alpha$
and the vector 
$\hat{\bm{n}}$ 
of unit normalization,
namely 
$\hat{\bm{n}} \cdot \hat{\bm{n}} = 1$. 
The normalization of the isospin generators is such that we restrict the angle 
$\alpha$ 
to
$0 \leq \alpha < 2 \pi$. 
The $\d$-regime Hamiltonian is determined from 
Eq.~\eqref{eq:Ham}, 
with the Laplace-Beltrami operator, 
$\mathfrak{D}^2$, 
having the general form
\begin{equation}
\mathfrak{D}^2
=
\frac{1}{\sqrt{g}}
\partial_{a}
\,
\sqrt{g}
\,
g^{ab}
\partial_{b}
,\end{equation}
where 
$g_{ab}$ 
as the induced metric on the coset manifold, 
which is given by
\begin{equation}
g_{ab}
=
\frac{1}{2}
\tr 
[\partial_a U^\dagger \partial_b U]
,\end{equation}
and satisfies
$g^{ab} g_{bc} = \delta^a {}_c$.
In the Laplace-Beltrami operator, 
$g$ 
is the determinant of the metric
$g_{ab}$.

For the particular case of the coset 
$U$
parametrized by 
Eq.~\eqref{eq:coset2}, 
we arrive at the 
$\d$-regime
Hamiltonian
\begin{equation}
\cH 
=
-
\frac{1}{\sin^2 \alpha}
\left(
\frac{\partial}{\partial \alpha}
\sin^2  \alpha
\frac{\partial}{\partial \alpha}
+ 
\bm{L}^2
\right)
- 
2 \mu^2
\cos \alpha
,\end{equation}
upon combining the Laplace-Beltrami operator on the manifold described by 
$U$ 
with the quark mass term of the chiral Lagrangian density. 
The quark mass term introduces breaking of the 
$SU(2)_L \times SU(2)_R$
symmetry of the Hamiltonian. 
The vector subgroup
$SU(2)_V$, 
however, 
remains in tact. 
Consequently the spectrum is described by irreducible representations of 
$SU(2)_V$, 
which are characterized by the isospin quantum number 
$\ell$. 
This quantum number appears in the eigenvalues of 
$\bm{L}^2$ 
in the familiar way, 
$\ell ( \ell + 1)$. 
On account of isospin symmetry,  
eigenfunctions of this 
$\d$-regime
Hamiltonian are of the form
\begin{equation}
\Psi_{n\ell m} (U)
=
\Psi_{n\ell}(\alpha)
\,
Y_{\ell m}(\theta,\phi)
.\end{equation}
The 
$Y_{\ell m} (\theta, \phi)$ 
are,
of course,  
the familiar spherical harmonics which are the characteristic functions for the isospin states 
$| \ell, m \rangle$. 
Thus, 
the crux of our task ahead lies in finding the spectrum of 
$\cH$
by determining the eigenfunctions
$\Psi_{n\ell}(\alpha)$.

To solve for these eigenfunctions, 
it is useful to cast the differential equation for the energy eigenvalues in a reduced form by making the replacement, 
$\Psi_{n\ell}(\alpha) = \psi_{n\ell}(\alpha) / \sin \alpha$, 
in which 
$\psi_{n \ell} (\alpha)$
is reminiscent of the reduced radial wavefunction from quantum mechanics. 
As a result, 
the energy eigenvalues, 
$E_{n \ell}$,
are determined by solving the differential equation
\begin{equation} \label{eq:ReducedHam}
\left[
-
\frac{d^2}{d\alpha^2}
-
1
+
\frac{\ell (\ell +1)}{\sin^2\alpha}
-
2\mu^2\cos\alpha
\right]
\psi_{n \ell} (\alpha)
=
E_{n \ell} \,
\psi_{n \ell} (\alpha)
.\end{equation}
In accordance with the 
$U(1)_L \times U(1)_R$ 
subgroup considered in 
Sec.~\ref{U1}, 
the equivalence of 
$\alpha$ 
modulo 
$2\pi$ 
leads to periodicity of the wavefunction in 
$\alpha$; 
however, 
there are further constraints imposed due to the double cover of 
$SO(3)$,
and the introduction of the reduced wavefunction,
$\psi_{n \ell} (\alpha)$. 
We can utilize the double cover to relate the wavefunction 
$\psi_{n \ell} (\alpha)$
on the interval
$0 \leq \alpha \leq \pi$
to that on 
$\pi < \alpha < 2 \pi$, 
namely 
$\psi_{n \ell} (\alpha + \pi ) = \psi_{n \ell} ( \pi - \alpha)$. 
As a result, 
we need only determine the wavefunction for 
$0 \leq \alpha \leq \pi$;
hence, 
the global properties of the manifold, 
$SU(2) = SO(3) \times  \mathbb{Z}_2$,
do not affect the spectrum. 
Finally the introduction of the reduced wavefunction gives us boundary conditions. 
As the original wavefunction 
$\Psi_{n \ell} (\alpha)$
must be finite to guarantee a normalizable solution, 
we require that 
$\psi_{n \ell}(0) = \psi_{n \ell} (\pi ) =  0$. 
Enforcing these conditions then determines a discrete set of energy eigenvalues,
$E_{n \ell}$.

\subsection{Determining the Spectrum}

To determine the spectrum via Eq.~\eqref{eq:ReducedHam}, 
we first specialize to the case of iso-singlet states, 
$\ell = 0$. 
With vanishing isospin, 
the eigenvalue problem is solved in terms of the odd Mathieu functions in the interval 
$0 \leq \alpha \leq \pi$, 
specifically of the form
\begin{equation}
\psi_{n0} (\alpha) 
=
N \texttt{se}_{2n} ( \frac{\alpha}{2},-4\mu^2)
,\end{equation}
where 
$n = 1, 2, \cdots$.  
Notice that the even Mathieu function solutions are disallowed by the imposition of boundary conditions, 
namely 
$\texttt{ce}_{2n} ( 0,-4\mu^2)\neq0$.
The iso-singlet spectrum is therefore determined by the odd Mathieu characteristics
\begin{equation}
E_{n0}
=
\frac{1}{4}
\texttt{b}_{2n}(-4\mu^2)
-
1
\label{eq:ell0}
.\end{equation}
These analytic solutions will prove useful in checking our numerical results.

%
%
%
%
%
%
\begin{figure}
\epsfig{file=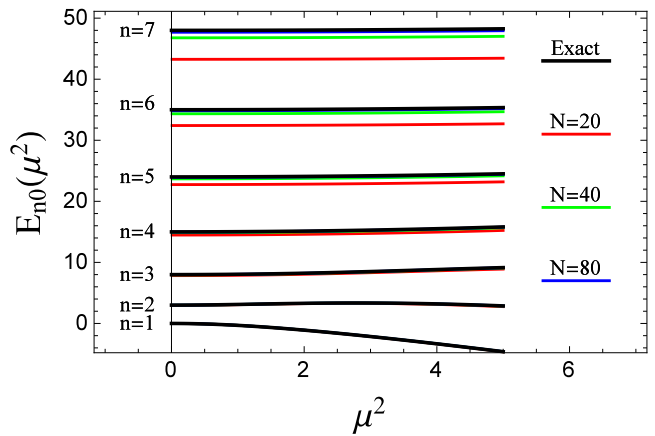,width=0.475\textwidth}
\caption{
The first seven energy eigenvalues of the 
$SU(2)_L \times SU(2)_R$ 
$\d$-regime Hamiltonian
with the vanishing isospin, 
$\ell=0$. 
The energies 
$E_{n 0} (\mu^2)$
are given analytically in Eq.~\eqref{eq:ell0}. 
Also depicted are numerical solutions to these energies for various values of the number of mesh points 
$N$. 
Values obtained for 
$N = 80$
are nearly identical to the analytic result. 
\label{f:convergenceSU2}
}
\end{figure}
%
%
%
%

In order to numerically solve for the eigenvalues with arbitrary isospin 
$\ell$, 
we cast the differential equation, 
Eq.~\eqref{eq:ReducedHam}, 
into a matrix eigenvalue equation. 
To accomplish this, 
we discretize the variable 
$\alpha$ 
into 
$N$ 
mesh points of even spacing 
$\Delta \alpha$ 
over the whole angular interval  
$\pi$. 
As a result, 
$\alpha$
takes on the discrete values
$\alpha_j= j \, \Delta \alpha$, 
with  
$\Delta \alpha = \pi / N$.
On the mesh, 
the second derivative is replaced by the finite-difference approximation
\begin{equation}
\frac{d^2 \psi_{n \ell} (\alpha_j) }{d\alpha^2} 
\to
\frac{\psi_{n \ell}(\alpha_{j+1})
-
2\psi_{n \ell}(\alpha_j)
+
\psi_{n \ell}(\alpha_{j-1})}
{\Delta \a^2}
,\end{equation}
which is valid up to terms of 
$O( \D \a^2)$. 
Using the notation 
$\psi_{n\ell} (\alpha_j ) \equiv (\psi_{n \ell})_j$
for the eigenvectors, 
the finite-difference approximation to the differential equation takes on the form 
\begin{equation}
\sum_{k=0}^N
\left[ 
\mathbb{M}_{\ell} (\mu^2)
\right]_{jk} (\psi_{n \ell})_k
= 
E_{n \ell} \, 
(\psi_{n \ell})_j 
,\end{equation}
where the matrix 
$\mathbb{M}_\ell (\mu^2)$
has the form
$\mathbb{M}_\ell (\mu^2) 
= 
\mathbb{T} + \mathbb{V}_\ell(\mu^2)$, 
with the kinetic term of the differential equation giving rise to the matrix
\begin{equation}
\mathbb{T}_{jk}
= 
- \frac{1}{\Delta \alpha^2}
\left[
\delta_{j,k+1}
- 
2 \delta_{jk}
+
\delta_{j,k-1}
\right]
\label{eq:Kin}
,\end{equation}
and the matrix potential emerging in diagonal form
\begin{equation}
[\mathbb{V}_\ell (\mu^2)]_{jk} 
= 
\delta_{jk}
\left[
-1 
+ 
\frac{\ell ( \ell+1)}{\sin^2 \a_j}
-
2 \mu^2 \cos \a_j
\right]
.\end{equation} 
To enforce Dirichlet boundary conditions at 
$\alpha = 0$
and 
$\alpha = \pi$, 
we define 
$\delta_{j,0} = \delta_{j,N} =0$
in Eq.~\eqref{eq:Kin}. 
Thus for a mesh of 
$N$ 
points, 
the matrix 
$\mathbb{M}_\ell (\mu^2)$
is a tridiagonal square matrix of dimension 
$N-1$. 
The low-lying eigenvalues can be found very efficiently using commercially available sparse-matrix techniques. 
Notice we must perform the matrix diagonalization for each isospin
$\ell$
and as a function of the parameter
$\mu^2$.

%
%
%
%
%
%
\begin{figure}
\epsfig{file=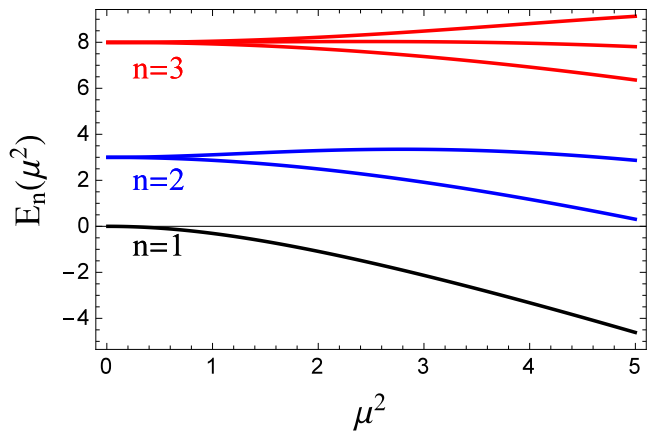,width=0.475\textwidth}
\caption{
Low-lying spectrum in the 
$\d$-regime of 
$SU(2)_L \times SU(2)_R$
chiral perturbation theory as a function of the mass parameter
$\mu^2 = (M_\pi L)^2 (F L)^4$. 
The ground-state (vacuum) energy is shown and
labeled by 
$n = 1$, 
as well as the higher levels with 
$n = 2$, $3$, $\cdots$. 
The splittings observed correspond to different  
isospin states,
with the smaller energies shown corresponding to larger 
$\ell$ 
values.
\label{f:specSU2}
}
\end{figure}
%
%
%
%

In order to check how well the finite-difference approximation is working, 
we compare our numerically determined results with the spectrum we obtained analytically for 
$\ell=0$. 
In Fig.~\ref{f:convergenceSU2}, 
we show how the numerical approximation converges to the exact solution as the number of mesh points 
$N$
increases. 
One should notice that on a fixed mesh of size
$N$,
the approximation is generally less accurate for the excited states of increasing 
$n$. 
We found the value
$N=80$
is more than sufficient to guarantee the accuracy of our results
(not just for the spectrum, but for the condensate calculated below).

Having checked the convergence of our numerical solution, 
we now present the spectrum for general isospin quantum number
$\ell$. 
The low-lying spectrum is plotted as a function of 
$\mu^2$
in 
Fig.~\ref{f:specSU2}. 
Notice that the principle quantum number 
$n$ 
has been defined to start at 
unity
in order to match with the  
$\ell=0$ 
case. 
As is well known, 
there is a high degree of degeneracy exhibited in the spectrum in the chiral limit, 
$\mu^2 = 0$.
Each energy level
$n$, 
has an 
$n^2$
degeneracy and contains states of differing isospin.
As 
$\mu^2$
increases, 
chiral symmetry is explicitly broken but isospin remains intact. 
We observe this symmetry breaking through mass splittings that depend upon the values of the isospin 
quantum number 
$\ell$. 
The lower-lying states correspond to those with larger isospin. 
For large 
$\mu^2$, 
a new degeneracy asymptotically appears dictated by the 
$SU(3)$
symmetry of the three-dimensional isotropic quantum harmonic oscillator. 
Quantum numbers of the low-lying multiplets are shown in 
Fig.~\ref{f:splits}, 
along with the evolution of the spectrum as a function of 
$\mu^2$. 
Excitation energies relative to the first excitation are plotted. 
These ratios are defined as

\begin{equation}
\Delta \cE_{n \ell}
=
\frac{E_{n \ell} -E_{10}}{E_{21} - E_{10}}
\label{eq:EEE}
.\end{equation}
The RR and SHO limits are investigated in detail for the mass gap below. 

\begin{figure}
\epsfig{file=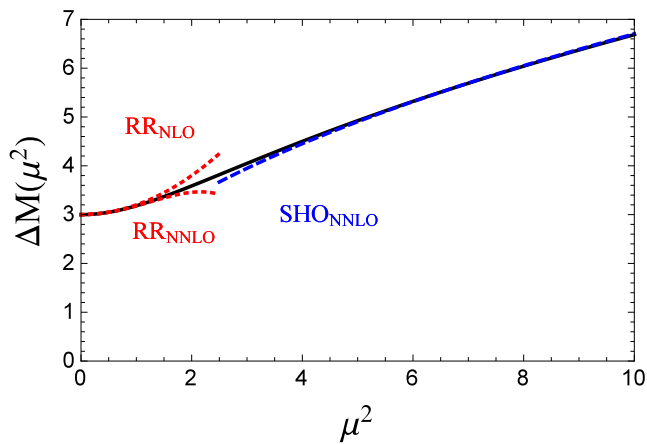,width=0.475\textwidth}
\caption{
Mass gap in the 
$\d$-regime of 
$SU(2)_L \times SU(2)_R$.
The gap 
$\Delta M$
given 
in 
Eq.~\eqref{eq:DMSU2},
is plotted as a function of 
$\mu^2$. 
Superimposed are results for the mass gap computed in perturbation theory for small 
$\mu^2$,
the 4-D rigid rotor 
(RR)
limit, 
and large 
$\mu^2$, 
the 3-D simple harmonic oscillator 
(SHO) 
limit.
\label{f:mgapSU2}
}
\end{figure}

\begin{widetext}

%
%
%
%
%
%
\begin{figure}
\epsfig{file=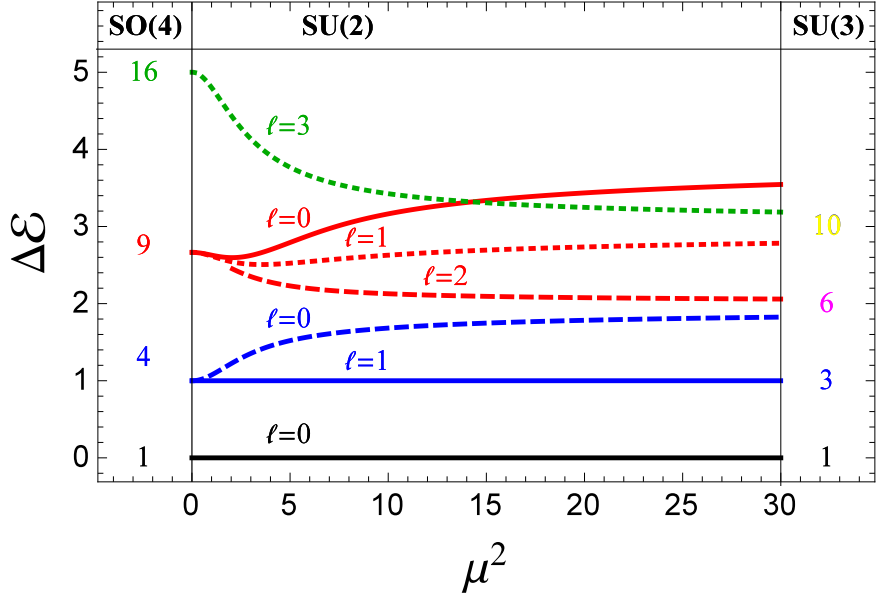,width=0.65\textwidth}
\caption{
Excitation energies 
$\Delta \cE_{n \ell}$
defined in 
Eq.~\eqref{eq:EEE}
shown as a function of 
$\mu^2$.
At vanishing 
$\mu^2$, 
the energy eigenstates fall into irreducible representations of the 
$SO(4)$ 
rigid rotor, 
while at asymptotically large
$\mu^2$, 
one encounters 
$SU(3)$
multiplets of the isotropic harmonic oscillator. 
The spectrum at intermediate values of 
$\mu^2$
maintains only 
$SU(2)_V$
isospin symmetry. 
The level crossing appearing in the plot occurs between states of differing isospin that cannot mix.
\label{f:splits}
}
\end{figure}
%
%
%
%

\subsection{$SU(2)_L\times SU(2)_R$ Mass Gap} 

Using the spectrum determined in the $\d$-regime, 
we compute the mass gap for 
$SU(2)_L\times SU(2)_R$
as a function of 
$\mu^2$. 
As in the 
$U(1)_L \times U(1)_R$
case, 
the first excited state in the spectrum corresponds to the pion; 
however, 
it is now a degenerate triplet with isospin
$\ell=1$. 
The
$m= \pm 1$, $0$ 
states
corresponding to the charged and neutral pions, 
respectively. 
The mass gap,
$\Delta M$, 
is thus given by
\begin{equation}
\Delta M
=
E_{21}
-
E_{10}
\label{eq:DMSU2}
,\end{equation}
where the first excited-state energy 
$E_{21}$
must be determined numerically. 
As in the previous case, 
the gap  
is non-vanishing in the chiral limit 
because there is no spontaneous symmetry breaking in finite volume
(and hence no Goldstone pions).  
Compared to the 
$U(1)_L \times U(1)_R$, 
the mass gap is larger in the present case. 
This fact can be attributed to the general flavor dependence of the mass gap in the $\d$-regime, 
namely%
~\cite{Leutwyler:1987ak}
\begin{equation}
\Delta M (\mu^2 = 0)
=
2  \,
\frac{N_f^2-1}{N_f}
\label{eq:gap}
,\end{equation}
where 
$N_f$
is the number of massless quark flavors, 
and we have rewritten the result in our dimensionless units. 
The mass gap determined as a function of 
$\mu^2$
is shown in 
Fig.~\ref{f:mgapSU2}. 
The behavior of the gap in limiting cases of 
$\mu$ 
can easily be discerned using 
Rayleigh-Schr\"odinger perturbation theory. 
The details are as follows.

\begin{itemize}
\item
4-D Rigid Rotor (RR).
At vanishing 
$\mu^2$, 
the effective Hamiltonian 
$\cH$
is that of a 
$SO(4)$
rigid rotor, 
with the eigenvalues 
$n^2-1$,
and wave functions 
\begin{equation}
\psi_{n\ell}(\alpha)
=
N
(\sin \alpha)^{\ell+1} \,
\texttt{C}_{n- \ell -1}^{\ell+1} \big( \cos \alpha \big)
,\end{equation} 
for 
$n = 1,2$, 
$\cdots$, 
and 
$l = 0$, $1$, $\cdots$,  $n-1$, 
where
$\texttt{C}^a_b(x)$ 
denotes Gegenbauer polynomials.%
\footnote{ 
While the matrix elements required in perturbation theory about the RR and SHO limits can both be evaluated using algebraic means, 
we found the explicit form of the eigenfunctions useful for additional checks on our numerical solutions, 
in particular for those states of non-singlet isospin, 
$\ell \neq 0$. }
Treating the quark mass term as a perturbation leads to the following selection rules: 
$\D n = \pm 1$  
and 
$\D \ell =0$.
All other matrix elements of the perturbation vanish.  
As a result,  
odd-order perturbative corrections vanish, 
and those higher-order terms which include them. 
The perturbative expansion of the mass gap is found to be
\begin{equation}
\D M^\text{RR}
=
3
+
\frac{1}{5}
\mu^4 
-
\frac{193}{9000}
\mu^8
+ 
\cO(\mu^{12})
.\end{equation}
The leading-order term is that in Eq.~\eqref{eq:gap}, 
while the 
$\mu^4$
term we deem next-to-leading order 
(NLO),
arises in second-order perturbation theory, 
and agrees with the result found in%
~\cite{Leutwyler:1987ak}. 
The 
$\mu^8$
term we deem next-to-next-to-leading order
(NNLO)
and we have determined it using fourth-order perturbation theory. 
\item
3-D Simple Harmonic Oscillator (SHO). 
In the opposite limit of large 
$\mu^2$, 
the potential term in 
$\cH$
dominates and freezes the angle 
$\alpha$ 
near zero.  
Fluctuations about this value lead to an harmonic approximation,
and the Hamiltonian becomes that of the three-dimensional isotropic harmonic oscillator. 
In this limit, 
the reduced wavefunction is given by 
\begin{equation}
\psi_{n \ell} (\alpha)
= 
N\alpha^{\ell +1} e^{- \frac{1}{2} \mu \alpha^2}
\texttt{L}_{\frac{1}{2} ( n - \ell - 1)}^{\ell+\frac{1}{2}}(\mu \alpha^2)
,\end{equation} 
where 
$N$ is the normalization constant, 
and 
$\texttt{L}_b^{a}(x)$ are the generalized Laguerre polynomials. 
The energy levels are given in the SHO approximation by 
\begin{eqnarray}
E_{n \ell}
&=&
-
2\mu^2
-
1
+ 
\frac{\ell(\ell+1)}{3} 
+
\left(
2 n+1
\right)
\sqrt{\mu^2 + \frac{\ell(\ell+1)}{15}}
\label{eq:SHOE}
,\end{eqnarray} 
where 
$n=1$,
$2$, 
$\cdots$. 
Notice the principle quantum number 
$n$
has been defined to start at unity to be consistent throughout 
Sec.~\ref{SU2}. 
The allowed values of 
$\ell$ 
are:
$\ell = 0$, $2$, $\cdots$, $n-1$
for $n$ odd, 
and 
$\ell = 1$, $3$, $\cdots$, $n-1$
for $n$ even.
The mass gap calculated in the SHO approximation is
\begin{equation}
\D M^\text{SHO}
=
2 \mu
+
\frac{1}{4}
+
\frac{13}{32\mu}
+ 
\cO(\mu^{-2})
,\end{equation}
where the first term is merely the difference in oscillator quanta. 
The second term (NLO) includes terms that arise in first-order perturbation theory,
and constant terms in the oscillator energy, 
Eq.~\eqref{eq:SHOE}. 
The third term (NNLO) comes from expanding Eq.~\eqref{eq:SHOE} to 
$\mu^{-1}$
order, 
as well as terms one order higher in perturbation theory. 
Reinstating the dimensionful parameters, 
we appropriately recover the pion mass from the mass gap in the infinite volume limit, 
namely
$\frac{1}{2 F^2 L^3} \Delta M^{\text{SHO}}  \overset{L \to \infty}{=} M_\pi$.

\end{itemize}

\end{widetext}

%
%
%
%
%
%
%
%
%
%

Referring back to 
Fig.~\ref{f:mgapSU2}, 
we see that the 
SHO
approximation appears to work well for the mass gap down to values of 
$\mu^2 \gtrsim 4$. 
In the opposite limit, 
we see that the 
RR
approximation remains under perturbative control up to values of 
$\mu^2 \lesssim \frac{3}{2}$, 
which, 
following the reasoning of the 
$U(1)_L \times U(1)_R$ case, 
corresponds to a restriction on the pion mass of 
$M_\pi \lesssim 110 \, \texttt{MeV}$
using a length of 
$L = 2.2 \, \texttt{fm}$.

\subsection{$SU(2)_L \times SU(2)_R$ Chiral Condensate}

%
%
%
%
%
%
\begin{figure}
\epsfig{file=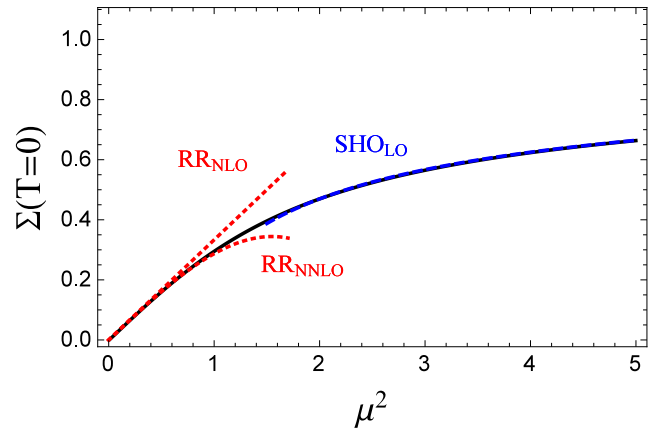,width=0.475\textwidth}
\caption{
Chiral condensate in the 
$\d$-regime 
of 
$SU(2)_L \times SU(2)_R$. 
Shown is the ratio of the chiral condensate at finite volume to that of infinite volume at zero temperature,
$T = 0$. 
Superimposed as checks are the RR approximation up to both NLO and NNLO accuracy, 
and the SHO approximation to LO accuracy.
\label{f:sig0SU2}
}
\end{figure}
%
%
%
%

We now utilize our solution for the spectrum of 
$SU(2)_L \times SU(2)_R$
chiral perturbation theory in the 
$\d$-regime to calculate the chiral condensate. 
At vanishing temperature, 
$T=0$, 
we can determine the condensate analytically.  
The sole occupied state is the iso-singlet ground state,  
and thus the ratio of finite to infinite volume condensates is given by
\begin{equation}
\Sigma(T=0)
=
- \frac{1}{8} 
\frac{\partial
\texttt{b}_2 ( - 4 \mu^2)}{\partial \mu^2}
\label{eq:cond2}
.\end{equation}
The behavior of the chiral condensate is depicted in 
Fig.~\ref{f:sig0SU2}. 
As expected, 
the condensate identically vanishes when 
$\mu^2 = 0$
on account of the absence of spontaneous symmetry breaking at finite volume. 
The introduction of a quark mass brings with it the explicit symmetry breaking
that appears in the potential of the $\d$-regime Hamiltonian.

Also shown in the figure are the limiting cases of 
$\mu^2$ 
dependence, 
between which the condensate determined from 
Eq.~\eqref{eq:cond2} 
nicely interpolates. 
The behavior in the limiting cases is determined as follows. 
About the RR limit we have, 
up to NNLO accuracy,
\begin{equation}
\Sigma^\text{RR}(T=0)
=
\frac{1}{3} \mu^2 
- 
\frac{5}{108} \mu^6
+ 
\cO(\mu^{10})
.\end{equation}
In contrast to the 
$U(1)_L \times U(1)_R$ 
condensate the RR approximation continues to perform well up to values of 
$\mu^2 \lesssim 1$. 
The RR approximation for the condensate is seen to be only a little less effective than the same approximation for the mass gap. 
This loss of effectiveness occurs because the zeroth-order term in the expansion of the condensate vanishes. 
The upper bound on 
$\mu^2$ 
for the condensate in the RR approximation 
corresponds to a pion mass satisfying 
$M_\pi \lesssim 90 \, \texttt{MeV}$,
for 
$L = 2.2 \, \texttt{fm}$. 
Taking up the opposite limit, 
$M_\pi L \gtrsim 1$, 
we have the SHO approximation.  
It works nearly as good as that in the 
$U(1)_L \times U(1)_R$
case. 
The distinction of keeping LO and NLO terms in the energy is moot because the NLO term is constant, 
and does not survive differentiation with respect to 
$\mu^2$.
For the SHO condensate, 
we have
\begin{equation}
\Sigma^\text{SHO}(T=0)
=
1
- 
\frac{3}{4} 
\mu^{-1}
+ 
\cO(\mu^{-2})
.\end{equation}
The condensate approaches the infinite volume limit by the same power of the volume,
namely 
$L^{-3}$, 
as in the 
$U(1)_L \times U(1)_R$ 
case; 
however, 
the differing numerical factor of 
$N_f^2 - 1=3$ 
makes it a slower approach.
As one nears the infinite volume limit, 
of course, 
one must add contributions from the non-zero modes. 
Summing these contributions produces exponentially small finite-volume effects.

%
%
%
%
%
%
\begin{figure}
\epsfig{file=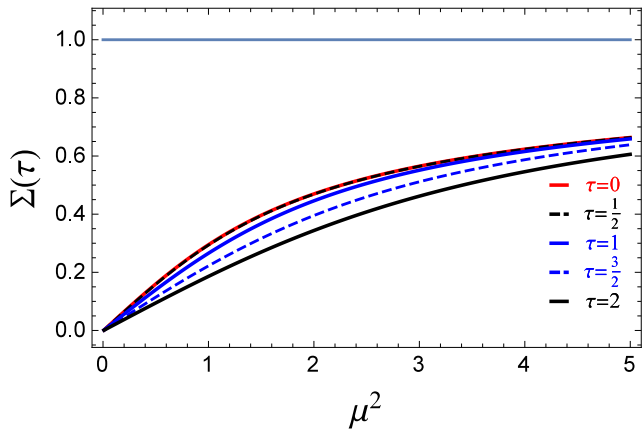,width=0.475\textwidth}
\caption{
Chiral condensate in the 
$\d$-regime 
of 
$SU(2)_L \times SU(2)_R$
as a function of mass 
$\mu^2$
and temperature. 
Shown is the ratio of the chiral condensate at finite volume to that in infinite volume for several values of the temperature, 
$\tau = 2 F^2 L^3 T$.
\label{f:sigTSU2}
}
\end{figure}
%
%
%
%

Increasing the temperature requires contributions to the condensate from excited states. 
We consider low temperatures for which truncation of the partition function is a good approximation. 
To this end, 
we use the same temperature range employed in the 
$U(1)_L \times U(1)_R$ 
case, 
and include states up to and including those corresponding to 
$n=7$. 
In order to compute the condensate, 
we account for the isospin degeneracy with proper factors of 
$g_\ell = 2\ell+1$, 
and we include in total
$28$ 
distinct states specified by 
$1\leq n \leq 7$ 
with 
$0\leq \ell \leq n$. 
Beyond this number of states, 
we see diminishing returns in accuracy for the highest temperature considered. 
The finite temperature behavior of the chiral condensate is shown in 
Fig.~\ref{f:sigTSU2}.
Comparing to 
$U(1)_L \times U(1)_R$, 
we see the condensate melts more slowly, 
but is also not as frozen to begin with.

\begin{figure}
\epsfig{file=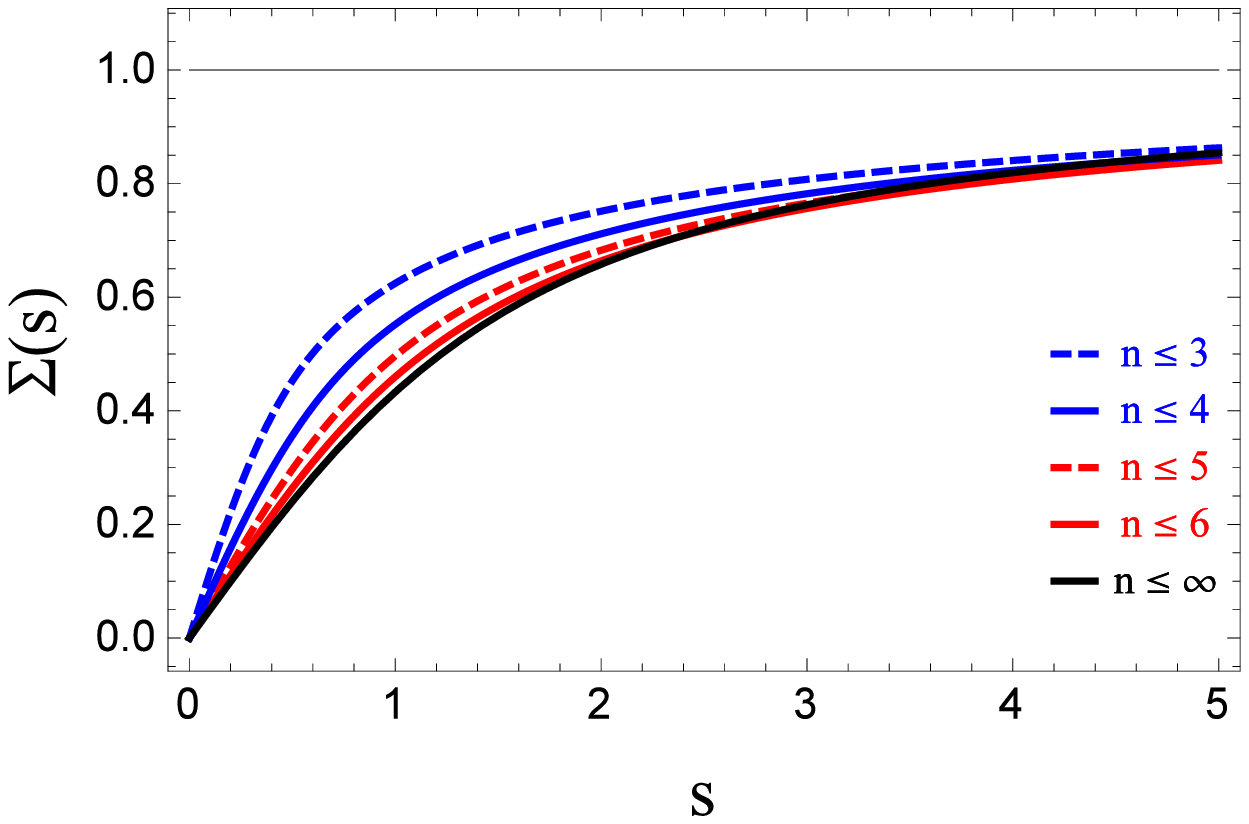,width=0.475\textwidth}
\caption{
Chiral condensate comparison between the 
$\d$-regime 
and
$\e$-regime
of 
$SU(2)_L \times SU(2)_R$
as a function of 
$s = \frac{1}{2} (M_\pi L)^2(FL)^2$. 
The convergence of the $\d$-regime condensate to that of the $\e$-regime is shown as a function of the number of energy eigenstates included in the partition function. 
For each $n$ value, 
all allowed $\ell$ states are included. 
We set 
$FL=\sqrt{8}$, 
which corresponds to a temperature of 
$T\sim 30 \, \texttt{MeV}$.
\label{f:sigepsSU2}
}
\end{figure}

To compare our computation of the chiral condensate in the 
$\d$-regime 
with that of the 
$\e$-regime, 
we must again set the temperature equal to inverse length, 
$T = L^{-1}$, 
which leads us to the additional condition
$M_\pi \beta \ll 1$. 
In the $\e$-regime, 
only the zero \emph{four}-momentum mode survives to leading order; 
therefore, 
there is no kinetic term in the effective chiral Lagrangian density. 
The partition function is then obtained by integration over the coset manifold
(which appears as a trace over the group coordinates), 
rather than by summation over energy eigenstates. 
The partition function in the $\e$-regime is thus given by%
~\cite{Gasser:1986vb}    
\begin{equation}
Z 
= 
\tr \left( e^{ -  \b H} \right)
=
\frac{1}{2\pi} \int_0^{2 \pi} d \a \, \sin^2 \a \,  e^{ 2 s \cos \a}
,\end{equation}
where the scaling variable 
$s = \frac{1}{2}(M_\pi L)^2(FL)^2$,
and satisfies
$s = \mu^2/ \tau$
when
$T = L^{-1}$. 
The chiral condensate ratio can be determined from applying 
Eq.~\eqref{eq:chi} 
to this partition function, 
and produces
\begin{equation}
\Sigma (s)
= 
\frac{d}{ds} \log \frac{\texttt{I}_1(2 s)}{2s}
,\end{equation}
where
$\texttt{I}_1 (x)$ 
is a modified Bessel function. 
In merging the two regimes, 
we seek to achieve a similar depiction as in the 
$U(1)_L \times U(1)_R$ 
case. 
For this purpose, 
we choose the value of 
$F L = \sqrt{8}$
which satisfies the condition 
$F L \gtrsim 1$, 
and corresponds to a temperature of 
$T \sim 30\, \texttt{MeV}$.
The convergence with the number of states of the $\d$-regime 
condensate to that of the $\e$-regime is depicted in 
Fig.~\ref{f:sigepsSU2}.
Compared to the 
$U(1)_L \times U(1)_R$
case, 
further excited states must be included at a lower temperature in order to have the two regimes meet.

%
%
%
%
%
%
%
%
%
%

\section{Summary} \label{summy}

In the above presentation,  
we investigate the
$\delta$-regime of chiral perturbation theory, 
which was first introduced and studied in~\cite{Leutwyler:1987ak}. 
This regime of low-energy QCD emerges in a box of finite spatial volume,
but with an adjustable temperature. 
Specifically pertinent is the requirement that the Compton wavelength of pions be larger than the length of the box, 
$1/M_\pi \gg L$. 
This condition necessitates non-perturbative treatment of the spatial zero modes of the Goldstone pions. 
The universal dynamics of these zero modes is governed by quantum mechanics on the coset manifold.
Details of the 
$\d$-regime power counting are reviewed in 
Sec.~\ref{PC}.  
Unlike the 
$\e$-regime, 
a novel feature of the 
$\d$-regime 
is the ability to explore the thermal behavior of the theory, 
provided the temperature remains in the low-energy regime, 
$M_\pi / T \lesssim 1$.

We investigate two scenarios of chiral symmetry relevant in low-energy QCD, 
namely that of 
$SU(2)_L \times SU(2)_R$,
and its
$U(1)_L \times U(1)_R$ 
subgroup.  
The simpler case of the 
$U(1)_L \times U(1)_R$ 
subgroup is considered first in 
Sec.~\ref{U1}. 
The energy spectrum is determined in terms of the well-known Mathieu functions, 
see 
Eq.~\eqref{eq:energ}. 
From the spectrum,  
we obtain the mass gap which corresponds to the mass of the neutral pion in finite volume. 
Unlike previous studies, 
we explore the pion mass and volume dependence by treating the quark mass according to the
$\d$-regime power counting. 
The spectrum and other derived quantities depend on the dimensionless parameter 
$\mu^2$, 
which is given by 
$\mu^2 = (M_\pi L)^2 (F L)^4$. 
To confirm our results, 
we consider two limiting cases for the values of  
$\mu^2$. 
In the large 
$\mu^2$ 
limit, 
the SHO approximation emerges, 
works quite well, 
and produces the pion mass in the infinite volume limit. 
In the small 
$\mu^2$ 
limit, 
the RR approximation emerges. 
We find that the success of perturbation theory about the RR limit requires quark masses that are lighter than physical. 
The chiral condensate is also computed and exhibits the expected behavior as a function of 
$\mu^2$
and temperature 
$T$. 
Our results nicely illustrate that melting of the condensate results from greater statistical weight attached to the excited states, 
which exhibit symmetry restoration.
For this
$U(1)_L \times U(1)_R$
case, 
symmetry restoration in the spectrum is exhibited by parity doubling. 
A stringent test of our results is provided by raising the temperature to meet up with analytical expectations from the 
$\e$-regime. 
As the number of states in the 
$\d$-regime
partition function is increased, 
the chiral condensate converges quickly to that of the 
$\e$-regime.

The case of 
$SU(2)_L \times SU(2)_R$
symmetry is considered in 
Sec.~\ref{SU2}. 
Unlike the previous case, 
an analytic solution for the entire spectrum is not known, 
and numerical methods are utilized to determine the energy eigenstates of non-zero isospin. 
Using a finite-difference approximation, 
the Hamiltonian is easily diagonalized using sparse-matrix techniques. 
As a result, 
properties in the 
$\d$-regime 
are determined as a function of the parameter 
$\mu^2$. 
The analytically soluble iso-singlet states provide a useful check on the numerical solution; 
and, 
on a fixed mesh, 
the accuracy decreases with increasing energy,
as expected. 
Many of the conclusions reached for the 
$U(1)_L \times U(1)_R$
case are mirrored in the
$SU(2)_L \times SU(2)_R$
case. 
In contrast, 
perturbation theory about the RR limit appears to work better for the mass gap, 
although slightly less-that-physical pion masses are required to approach this limit in practice. 
The spectrum shows rich behavior as a function of 
$\mu^2$. 
Energy eigenstates fall into  
$SU(2)_L \times SU(2)_R$ 
multiplets for small 
$\mu^2$, 
$SU(2)_V$
multiplets for intermediate 
$\mu^2$, 
and 
$SU(3)$
multiplets for large 
$\mu^2$. 
The chiral condensate exhibits the expected behavior at small temperatures, 
however, 
more numerous excited states are required to achieve convergence of the partition function compared to the 
$U(1)_L \times U(1)_R$
case. 
Finally, 
we use the numerically determined condensate in the $\d$-regime to reproduce the analytically known
condensate of the 
$\e$-regime. 
The observed melting of the chiral condensate seen in the 
$\e$-regime 
requires the inclusion of 
$28$ 
states in the 
$\d$-regime 
partition function at a modest temperature of 
$T = 30 \, \texttt{MeV}$.

We show that numerical solution of the 
$\d$-regime Hamiltonian provides useful insight into the properties of 
low-energy QCD. 
In particular, 
the rich symmetry properties of the spectrum are nicely illustrated by computing the 
quantum mechanical energy eigenvalues. 
It should prove rewarding to numerically explore the case of 
$SU(3)_L \times SU(3)_R$
symmetry, 
which is also of relevance for low-energy QCD. 
Other groups can similarly be explored in the 
$\d$-regime, 
especially those relevant for condensed matter systems with spontaneously broken symmetries. 
Finally confrontation with numerical data from lattice QCD computations is desirable. 
We intend to use our results to explore the extent to which existing computations enter the various finite-volume
regimes depicted in 
Fig.~\ref{f:realms}.

\begin{acknowledgments}
We gratefully acknowledge support for this work from the U.S.~National Science Foundation,
under Grant No.~PHY$15$-$15738$, 
and from grants from the Professional Staff Congress of The CUNY. 
The work of BCT is additionally supported by a joint 
The City College of New York--RIKEN/Brookhaven Research Center fellowship.
We thank T.~Mehen and M.~Nomura for fruitful discussions and involvement during early stages of this work. 
\end{acknowledgments}

\appendix

\newpage

\bibliography{bibly}

\end{document}